# Predicting Stock Price of Construction Companies using Weighted Ensemble Learning

Xinyuan Song[1,a*]

[1]Department of statistics and data science, National University of Singapore, Lower Kent Ridge Road 119077, Singapore
[a]Email: songxinyuan@pku.org.cn
[*]Corresponding Author

**Abstract**

Modeling the behavior of stock price data has always been one of the challengeous applications of Artificial Intelligence (AI) and Machine Learning (ML) due to its high complexity and dependence on various conditions. Recent studies show that this will be difficult to do with just one learning model. The problem can be more complex for companies of construction section, due to the dependency of their behavior on more conditions. This study aims to provide a hybrid model for improving the accuracy of prediction for stock price index of companies in construction section. The contribution of this paper can be considered as follows: First, a combination of several prediction models is used to predict stock price, so that learning models can cover each other's error.  In this research, an ensemble model based on Artificial Neural Network (ANN), Gaussian Process Regression (GPR) and Classification and Regression Tree (CART) is presented for predicting stock price index. Second, the optimization technique is used to determine the effect of each learning model on the prediction result. For this purpose, first all three mentioned algorithms process the data simultaneously and perform the prediction operation. Then, using the Cuckoo Search (CS) algorithm, the output weight of each algorithm is determined as a coefficient. Finally, using the ensemble technique, these results are combined and the final output is generated through weighted averaging on optimal coefficients. The proposed system was implemented, and its efficiency was evaluated by real stock data of construction companies. The results showed that using CS optimization in the proposed ensemble system is highly effective in reducing prediction error. According to the results, the proposed system can predict the price index with an average accuracy of 96.4%, which shows reduction of at least 2.4% of prediction error compared to the previous methods. Comparing the evaluation results of the proposed system with similar algorithms, indicates that our model is more accurate and can be useful for predicting stock price index in real-world scenarios.

# 1. Introduction

Investing in the stock market can be considered as one of the most important parts of the economy in each country, and most of the capital around the world, is traded through stock markets. There is no doubt that the economy of any country is strongly affected by performance of its stock market [1]. During recent years, the financial markets around the world have faced significant fluctuations and uncertainties, so that the existing uncertainty about the return of investments has worried many financial analysts and investors [2]. As investors point out, uncertainty is considered as the most important factor in determining the actual price of financial assets [3, 4]. A large number of studies have been done about stock value forecasting using machine learning approaches; because, providing a proper approximation of the price of a stock for subsequent years is very important. This issue is of higher importance for companies in the construction section. Because, the high dependence of the asset value of these companies on various economic and political conditions causes the instability of their stock prices (especially in crisis conditions), which can lead to more concerns for investors. In addition, the activity pattern of construction companies in the stock market can be considered as a highly influential factor on the growth rate of developing countries, and for this reason, modeling their functional pattern can be effective in more accurate planning of governments.

Studying and analyzing the construction company's profitability data, can be effective in prediction of its stock price, but it may not lead to an efficient approach. On the other hand, simple regression methods cannot determine the amount of profitability [5]. Therefore, this paper presents an ensemble method for prediction of stock price index in construction companies. Studies on the application of ensemble learning for predicting stock price are not rich enough.

The contribution of this paper is twofold:

- First, a combination of several ensemble learning-based prediction models is used to predict stock price, so that learning models can cover each other's error. To this aim, the proposed system uses a set of prediction models including: Artificial Neural Network (ANN), Gaussian Process Regression (GPR) and Classification and Regression Tree (CART) to improve the prediction accuracy.
- Second, the optimization technique is used to determine the effect of each learning model on the prediction result. For this purpose, the proposed approach usesCuckoo Search (CS) optimization to rank the prediction algorithms. This optimization algorithm, by assigning a weight value to each learning model, determines how effective the prediction result of that model is in determining the final output.

Based on our studies, both of these cases have not been addressed in previous researches and they can be considered as innovative aspects of this research. The continuation of this paper is organized as follows: Related works are reviewed in section two, and the proposed approach is described in the section three. The implementation results are presented in the fourth section, and the findings are discussed in the fifth section.

## 2. Related Works

Predicting the future price of a financial asset is essential for investors, as they can reduce the risk of a decision by properly determining the future movement of an investment asset. Given the history of stock price changes, predicting its next movement has always been a challenge. These approaches include discovering the appropriate pattern of market data and providing an optimal mechanism for investment decisions [6]. Analysis and forecasting in the stock market has been studied using various methods such as machine learning and text analysis.

In [7], the combination of machine learning techniques and news analysis in Twitter data is used to predict stock price. Chang et al. [8], have developed Organizational Sustainability Modeling (OSM) that can quickly process a large amounts of financial data. Ariyo et al [9], have proposed an Auto-Regressive Integrated Moving Average (ARIMA)-based stock value prediction system and tested their system with data from the New York and the Nigeria Stock Exchange. They concluded that the ARIMA model has significant potential for short-term prediction of stock price. Although these econometric models are suitable for describing and evaluating the relationships between variables with statistical inference, they have limitations in analyzing the time series of financial data.

Lendasse et al [10], proposed a stock index prediction system based on an ANN with Radial Basis Function (RBF). They showed that their system could record nonlinear relationships in financial time series data. In [11], a Support Vector Regression (SVR) model has been used to predict the movement of stock price data. The time series used in this model provides the ability to predict at intervals of one day to one minute, which distinguishes this research from previous solutions.

During recent years, the application of deep learning models to predict the stock price in the market has received more attention. For example, the study conducted in [12], which examines three different deep learning models: Long Short-Term Memory (LSTM), Recurrent Neural Network (RNN) and Convolutional Neural Network (CNN) to predict the short-term value of stock price. This study reports the superiority of deep learning techniques over other learning techniques. The method presented in [13] also uses a deep learning model based on Phase Space Reconstruction (PSR) and LSTM network for short-term prediction of companies' stock price. In [14], the application of different regression techniques to predict the companies' stock price at weekly intervals is evaluated. This research shows that CNN can achieve higher accuracy in predicting stock price. The research presented in [15] has also evaluated the efficiency of deep learning techniques in comparison with other machine learning techniques. This research shows that some learning models such as ANNs and decision trees have a performance close to deep learning techniques.

Research in [16], considers a closed-loop supply chain by taking into account sustainability, resilience, robustness, and risk aversion. The authors, have suggested a two stage Mixed-Integer Linear Programming (MILP) model for the problem and the counterpart model is used to handle uncertainties. Researchers in [17], have focused on the problem of sustainable development, and investigated the time-cost-quality-energy-environment problem in executing construction projects using a robust approach. The authors have tried to take into account the sustainability pillars in scheduling projects and uncertainties in modeling them. They have applied robust Non-Linear Programming (NLP) approach to model the study problem which uses cost, quality, energy, and pollution level as objectives. In [18], a robust, resilient, risk-aware and sustainable closed-loop supply chain network model has been proposed which is resistant against demand fluctuation like COVID-19 pandemic. This study, uses a two-stage robust stochastic multi-objective programming model to formulate the problem. This model, uses objectives such as $CO_2$ emissions, costs and energy consumption.

Optimization, is a useful technique for solving a wide range of real-world problems. Many researches have considered optimization techniques as an approach for solving NP-Hard problems. In [19], the task of designing a sustainable closed-loop supply chain network for face masks has been modelled as a multi-objective optimization problem and two algorithms of: Non-Dominated Sorting Genetic Algorithm II (NSGA-II) and Multi-Objective Grey Wolf Optimization (MOGWO) have been implemented for solving the problem. Researchers in [20], have proposed an intelligent method for evaluating the performance of supply chain, which can be divided into several minor methods such as: hybrid Shapley value path analysis, balanced scorecard and Multimoora method.

Research in [21], provides an optimization model for improving the process of allocating and routing relief vehicles, at the time of crisis. Using covering tour mechanism, this method tries to minimize response time and defines the demand parameter as a fuzzy variable to better match with the uncertainty of the situations. Finally, Harmony Search (HS) algorithm has been used to solve the optimization problem. Researchers in [22], have introduced a HS-based model to optimally allocate and route temporary health centers at the time of crisis. This research introduces an improved version of HS algorithm, which demonstrates an improvement compared to the classic HS algorithm. In [23], meta-heuristic algorithms have been used to solve the problem of group scheduling in sequence-dependent parallel machines. In this research, first the problem has been modelled mathematically. Then, the Biogeography-based Optimization (BBO) algorithm has been applied to solve the problem.

Research in [24], studies the problem of optimizing product portfolio under return uncertainty conditions. This method, attempts to predict the future demand of products using an improved ANN which is optimized by root runner algorithm. In a similar research [25], the combination of machine learning and meta-heuristic algorithms for predicting future demand of dairy products has been investigated. Researchers in [26, 27], have evaluated the efficiency of hybrid algorithms in predicting wind speed. In these researches, the combination of ANN with several optimization algorithms have been examined. The results indicate that optimizing the learning models can be effective in improving the predicting performance.

Research in [28], has modelled the problem of renewable energy location as a multi-objective optimization problem. In this model, profit and energy have been considered as the goals of optimization which correspond to objectives of supplier and government; then, an improved augmented $\varepsilon$-constraint model has been used for solving it. Research in [29], have introduced a viable closed-loop supply chain network by considering robustness and risk as a circular economy. This model is effective in decreasing the cost function, energy consumption grow, time solution and allowed maximum energy. In [30], the problem of resource-constrained time–cost-quality-energy-environment tradeoff by considering blockchain technology, risk and robustness has been investigated. This research, has presented a real case study in healthcare section and utilized GAMS-CPLEX for solving it. Related works are summarized in Table 1.

**Table 1: Summary of the studied researches**

| Reference | Year | Research method | Advantage(s) | Disadvantage(s) |
|---|---|---|---|---|
| [7] | 2011 | stock price prediction by news analysis in Twitter | using high level and real-world features for prediction | high complexity and difficult to implement |
| [9] | 2014 | ARIMA-based stock value prediction | fully describing the relationships between variables with statistical inference | limitations in analyzing the time series of financial data |
| [10] | 2000 | stock index prediction system based on RBF ANN | usable for long-term and short-term predictions | high error rate, high computational complexity |
| [11] | 2018 | predicting the movement of stock price data by SVR | usable for short-term to mid-term predictions | probability of using unrelated features for prediction |
| [12] | 2017 | predict the short-term value of stock price by LSTM, RNN and CNN | high prediction accuracy | limited to short-term predictions, high computational complexity |
| [13] | 2020 | short-term stock price prediction by PSR and LSTM | high prediction accuracy | high computational complexity, limited experiments |
| [14] | 2020 | weekly prediction of stock price by CNN | covering short-term to mid-term prediction scenarios | using a large set of features which makes it hard to implement |
| [15] | 2020 | comparing the performance of ML and DL methods in prediction | considering a large number of methods in comparison | high error rate due to limitation on used feature |

| Reference | Year | Research method | Advantage(s) | Disadvantage(s) |
|---|---|---|---|---|
| Proposed Method | 2023 | predicting stock price of construction companies by weighted ensemble learning | high prediction accuracy, determining the effect of learners in ensemble model according to their performance dynamically | higher computation time during training phase |

Despite numerous researches in predicting stock price, the study of methods based on ensemble learning has received less attention; however, it has been theoretically proven that the use of ensemble models improves the performance of learning models compared to the case where these techniques are used individually. Accordingly, in this study, the application of ensemble technique in the analysis of stock data is studied and an efficient solution to predict stock price based on weighted ensemble learning will be presented.

## 2.1. CS Optimization

CS, is a search strategy that compared to its previous optimization techniques, has several benefits such as simplicity of operations and fast convergence. The cuckoo search mechanism starts by defining random values for each optimization variable and then using an iterative approach to improve them. The CS algorithm, follows three basic rules:

1. Each time an egg is placed in a nest by cuckoo agents, randomly.
2. At the end of each cycle, the best nests, containing eggs with the best fitness are saved and transferred to the next cycle. This means that the most suitable solutions will be present in the next iterations.
3. In each iteration, there are a certain number of available nests, and $P_a \in (0,1)$ is the probability of finding alien egg by the host.

The pseudo code of CS is presented in the following [31]:

| Algorithm 1: CS algorithm |
|---|
| 1- Define the fitness evaluation function as $f(x) = (x_1, \ldots, x_d)$<br>2. Initialize population of n nests, and place K random eggs (solutions) in each one.<br>3. While $(t < T)$ repeat this step:<br>   3-1- Use Levy algorithm to choose a random cuckoo such as $i$.<br>   3-2- Calculate the fitness of chosen cuckoo (using $f$).<br>   3-3- Select a nest such as $j$, randomly.<br>   3-4- If solutions of $j$ dominate $i$, then replace $i$ with the solutions in $j$.<br>   3-5- Find a ratio of $P_a$ nests with worst fitness and replace them.<br>   3-6- Store the fittest solutions for next cycle.<br>   3-7- Update the counter of algorithm iterations $(t)$.<br>4- End while. |

The CS algorithm starts with generating a population of random solutions (known as eggs) for the nests. Then, an iterative approach is used to improve the population.. During each cycle, a cuckoo is selected by Levy flight algorithm and its fitness if evaluated. Then a random nest (solution set) is selected and compared with the first choice. If members of the second chosen nest dominate the first selection, then the members of defeated solution set ate replaced by the winner. In the next step, a fraction of $P_a$ nests with worst fitness are abandoned, and a new set of random solutions are replacing them. The best set of solutions are maintained and transferred to next cycle. These steps are repeated for a predetermined number of iterations (T) [31]. In the method proposed in this paper, the cuckoo search algorithm is used to determine the optimal weight values of the learning models. In the following, we will describe how it works.

## 3. Proposed Method

This section is dedicated to describing the proposed system for predicting stock price of construction companies using weighted ensemble learning, in detail. The overall structure of ML-based stock price prediction systems consist of: data preparation, training and forecasting steps. The proposed method, using the ensemble learning technique, follows this structure in addition to an optimization step which tries to determine the optimal weight of ensemble learners. By weighting the output of the learning models, the effect coefficient of each model can be determined according to its training performance, which leads to increasing the overall accuracy of the ensemble system. For this reason, the optimization step is considered as an effective phase for increasing the prediction accuracy of the proposed method. Accordingly, the proposed system performs the prediction through the following steps:

1. Preparation of data by describing the time-based stock price series.
2. Training the learners in the proposed ensemble model using training samples. The proposed ensemble model includes three learners:
   - ANN.
   - GPR.
   - CART.
1. Weighting the learners of ensemble model based on the quality of training and using cuckoo search algorithm.
2. Determining the output of the proposed system by weighted averaging of the three algorithms used.

The prediction process using the proposed system is illustrated as a diagram in Figure (1). Also, the notations used in this paper are listed in Table (2).

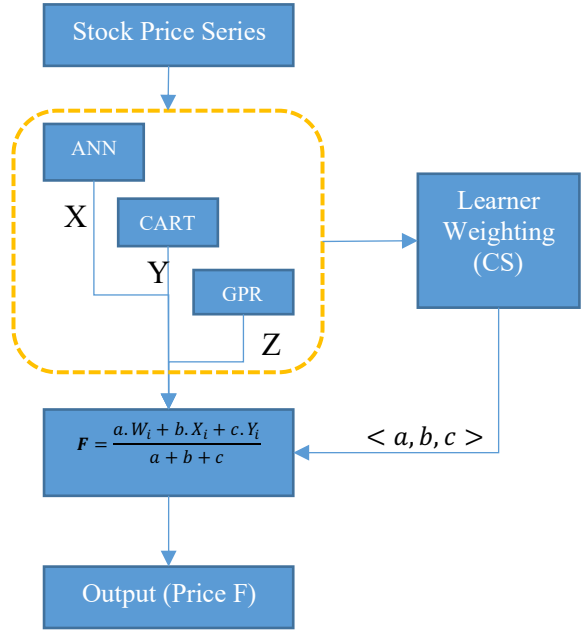

*Figure 1: The diagram of the prediction process using the proposed method*

**Table 2: the notations used in this paper**

| Symbol | Description |
|---|---|
| $a$ | Weight value of ANN in weighted ensemble system |
| $b$ | Weight value of CART in weighted ensemble system |
| $c$ | Weight value of GPR in weighted ensemble system |
| $s(i)$ | The $i$th element in the time series of stock final prices |
| $t$ | timeslot between each pair of elements in the sequence stock prices |
| $n$ | sequence size of stock prices |
| $C_r(x, y)$ | Correlation between two sequences of x and y |
| $\mu_x$ | Average of sequence x |
| $EMA_x$ | the moving average of $x$ days |
| $N_{gain}$ | Number of profitable days in a target period |
| $N_{lost}$ | Number of unprofitable days in a target period |
| $A_c$ | Average of final prices in construction section |
| $S_c$ | Standard deviation of final prices in construction section |
| $i(t)$ | The Gini index of attribute t |
| $p(w)$ | The probability of occurring $w$ in observations |

In the proposed method, the modelling of the time-based stock price data is done using the stock price history records and technical analysis indicators such as: Relative Strength Index (RSI) and Moving Average Convergence Divergence (MACD). This time-based data model is used as input to learning algorithms. It should be noted that the weighting of prediction algorithms is done only once and after the training operation. In the test phase of the proposed method, first the output value of each prediction algorithm is calculated separately. Then, by combining these values, the final output is determined based on the calculated weights. If for an input sample such as $i$, the ANN outputs $W_i$, the CART produce output value of $X_i$, and the GPR produce output $Y_i$, and also, these algorithms are weighted as: $a$, $b$ and $c$, respectively; then, the final output of the proposed method for each test sample is then calculated as follows:

$$F = \frac{a.W_i + b.X_i + c.Y_i}{a + b + c} \tag{1}$$

The above equation creates a linear combination of the output of the learning models and tries to minimize the prediction error by calculating a weighted average of learner's outputs. To achieve this goal, it is necessary to determine the optimal values for weights $a$, $b$ and $c$. In the proposed system, the CS optimization is used for determining the optimal values of these weights which will be described in Section 3-5. In the following, the computational steps of the proposed method is described in details.

### 3.1. Describing the time-based stock price series

Today, the parameters of stock analysis systems are collected using technologies such as information networks and databases. These solutions can collect market price information at different time intervals. In the proposed method, the values of final price of stocks will be used for prediction. Let's define the time series of stock final prices as $\{s(1), s(2), ..., s(N)\}$. Then, the following sequence can provide the input to understand the stock price in time-step $k$:

$$S(k) = [s(k - n \times t), ..., s(k - t), s(k)] \tag{2}$$

Where, $t$ refers to the timeslot between each pair of elements in the input sequence and $n$ describes the sequence size. For example, if we choose $n = 3$ and $t = 5$, then the sequence of Equation (2) for time $k$ will be described as $S(k) = [s(k - 15), s(k - 10), s(k - 5), s(k)]$. Thus, for modelling a stock value prediction system, in addition to the current stock price, the stock price at previous regular intervals should be contained in the input sequence. It should be noted that forecasting stock price may not lead to favorable results just by considering the history of price changes of that stock. That's why the proposed system uses the following features to construct a stock price data model:

*A) The correlation with the market and construction index*
Using these two features, we can describe the degree of concurrence of changes in the final price of the stock with the construction index and market index. For this purpose, the correlation between the market index/construction index and the final stock price in the last 14 days will be calculated using the following equation [32]:

$$C_r = \frac{\sum_{i=1}^{14}(x_i - \mu_x)(y_i - \mu_y)}{13} \qquad (3)$$

Where, $x_i$ indicates the final stock price on the $i$-th day and $\mu_x$ shows the average of final prices in the last 14 days. $Y_i$ also describes the value of the market index/construction index on the $i$-th day and $\mu_y$ is the average of the market index/construction index in the last 14 days.

*B) Market index value*

For a stock that has a high correlation with the market index, knowing the current value of the market index can be effective in predicting the value of that stock. For this reason, the value of the market index in the last 7 days will be used as another descriptive feature in the proposed data model.

*C) Average and standard deviation of price in construction section*

Companies of the construction section often respond to political issues, news and economic conditions, with a similar pattern. This situation is shown in Figure 2, which illustrates the final price of 15 companies in construction section for a 30 days period.

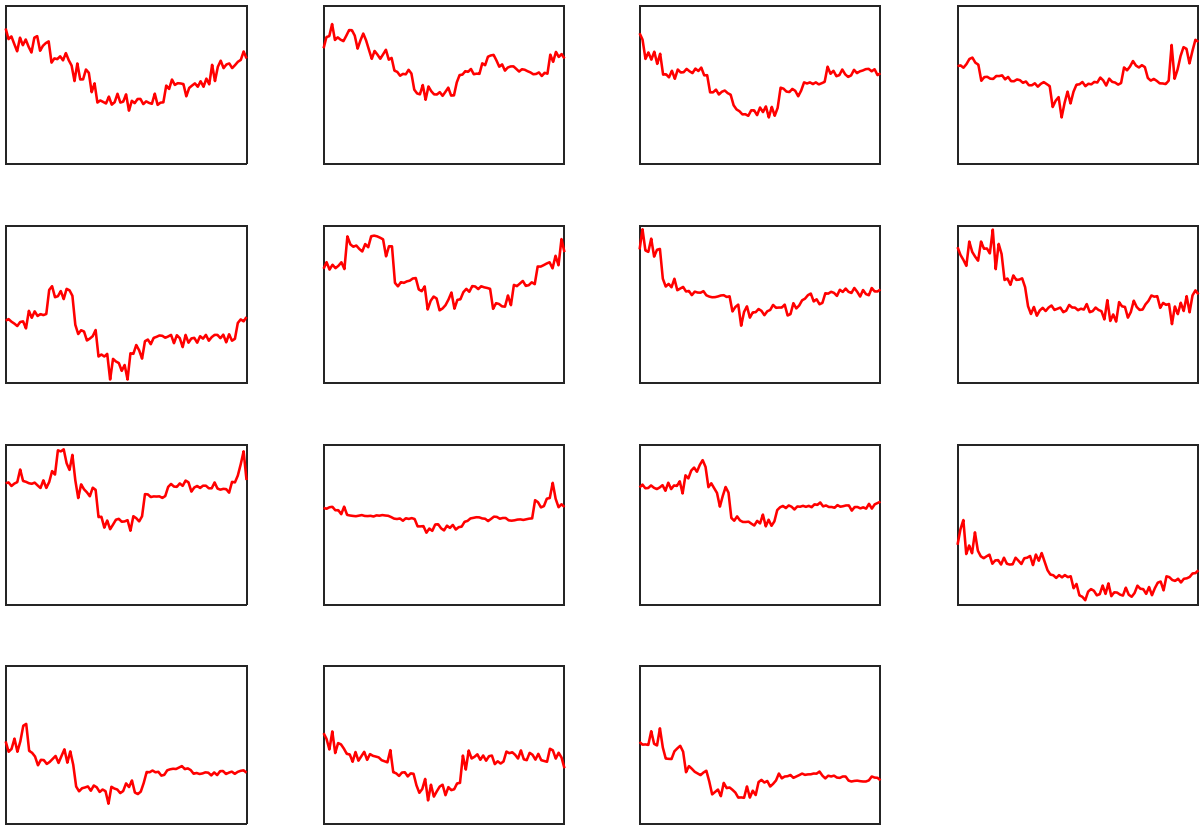

*Figure 2. The final price of 15 companies in construction section for a 30 days period*

The information shown in Figure 2 is related to the changes in the final stock price of construction companies during the Iran JCPOA negotiation period (July 2022). As shown in this figure, most companies have similar reaction patterns to the news of negotiations and political conditions of the country. For this reason, using the characteristics of the standard deviation and average of the value of all construction companies in the recent period can be effective in predicting the future status of a company more accurately.

*D) MACD indicator*

This indicator can be calculated based on the moving average. To calculate this indicator, the long-term moving average (26-day moving average) is subtracted from the short-term moving average (13-day moving average) as follows [33]:

$$M = EMA_{13} - EMA_{26} \qquad (4)$$

Where, $EMA_x$ represents the moving average of $x$ days.

*E) RSI indicator*

This indicator is calculated by considering the average profit and average loss over a period of 14 days [33]:

$$R = 100 - \frac{100}{1 + RS} \qquad (5)$$
$$RS = \frac{N_{gain}}{N_{lost}}$$

Where, $N_{gain}$ indicates the profitable days in the target period. Also, $N_{lost}$ indicates the number of unprofitable days in this period.

Considering the above features, Equation (2) is rewritten to describe the stock price model as follows:

$$S(k) = [C_r, C_c, I_r, A_c, S_c, M, R, s(k - n \times t), \dots, s(k - t), s(k)] \qquad (6)$$

Where, $C_r$ and $C_c$ represent the correlation with the market index and construction index, respectively. $I_r$, represent the value of the market index. $A_c$ and $S_c$, refer to Average and standard deviation of final prices in construction section, respectively. $M$ and $R$ also represent the calculated values for the MACD and RSI indicators, respectively. The proposed stock price prediction model tries to provide an appropriate estimate of stock price at time $k + 1$ using the input $s(k)$:

$$p(k + 1) = f(s(k)) \qquad (7)$$

In the above equation, *p* describes the predicted value (final price) of the stock. The goal of the proposed system is to provide a suitable approximation for function (*f*) using the weighted ensemble learning approach.

**3.2. Prediction by ANN**

One of the algorithms used to predict the price of stocks is an ANN with two hidden layers. This neural network has 10 neurons in its first, and 7 in its second hidden layer. The transfer function of these layers is of type logarithmic sigmoid. The size of input layer, is determined by the count of input attributes of each sample in Eq. (6) and also one neuron is placed in the output layer of this ANN. The proposed ANN structure is shown in Figure (3).

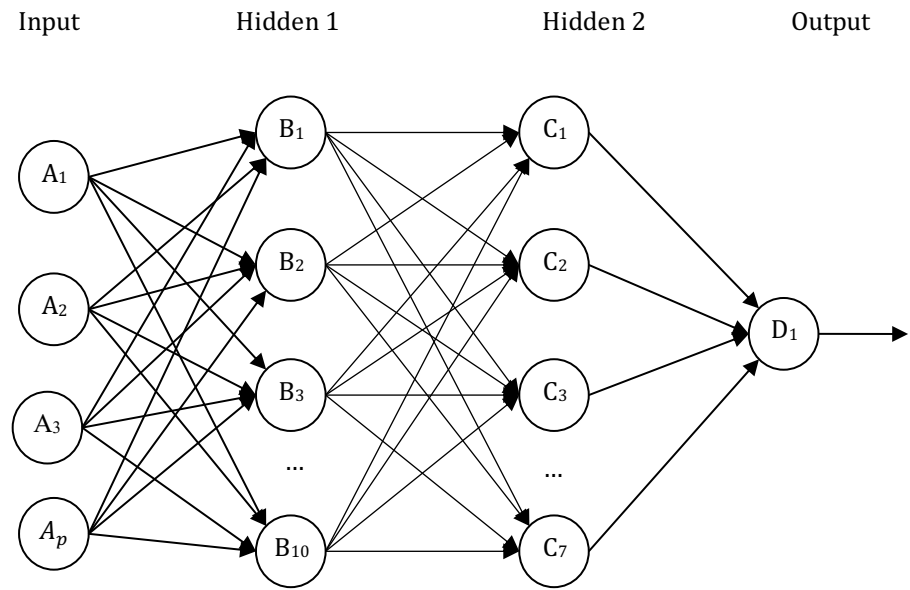

*Figure 3: The structure of ANN to predict stock price in proposed method*

The proposed ANN structure was trained using Levenberg-Marquardt (LM) backpropagation algorithm. The LM algorithm, trains ANN by minimizing the output absolute based on the Jacobean matrix. The training mechanism of this algorithm is fully described in [34]. Therefore, we will skip describing the computational details of this training algorithm.

**3.3. Prediction by CART**

This section presents a structure based on the CART to predict stock prices. The CART produces a decision tree that tries to predict or classify future observations. This method tries to reduce impurities in each category. A node is completely free of impurities when all the elements of a subset belong to a target category. In this tree, all divisions are in binary form; meaning that only two subgroups of each node will branch. The CART used Gini index as a measure of impurity [35]:

$$i(t) = 1 - \sum_{j=1}^{J} p^2(j|t) \tag{8}$$

In above equation, $p(j|t)$ is the estimated probability that $t$ belongs to class $j$, and $J$ specifies the number of classes. Suppose a branch divides the data $t$ into child nodes on the left $t_L$ and on the right $t_R$. Also, $p_L$ and $p_R$ are segments of $t$ at nodes $t_L$ and $t_R$, respectively. The CART creates a branch that can produce the maximum reduction in impurity $(i(t) - p_L i(t_L) - p_R i(t_R))$. It produces a sequence of sub-trees by producing a large tree instead of using stop laws, and then pruning the tree until only the root node remains. In the next step, cross-validation is used to estimate the classification cost of each sub-tree. Finally, a sub-tree with the lowest estimated cost is selected and the tree model is formed.

### 3.4. Prediction by GPR

The third model used for predicting stock price in the proposed system is the GPR, which is a Bayesian non-parametric approach for regression operations. GPR has several advantages; For example: it works well on small to medium-sized datasets and provides a degree of confidence in the estimated values in prediction [36].

Unlike many widely used supervised ML algorithms that try to learn the exact values of all parameters within a function, in Bayesian methods the probability distributions of values in parameters is deduced. For example, in a linear function as $y = wx + e$, The Bayesian method specifies the prior distribution of the $w$ parameter and the displacement probabilities based on the observed samples using the Bayesian rule and is described as follows [36]:

$$P(w|y, X) = \frac{P(y|X, w)p(w)}{p(y|X)} \Rightarrow posterior = likelihood \times \frac{prior}{marginal\ likelihood} \tag{9}$$

Where, $P(w|y, X)$ indicates the probability that the value of $w$, based on observations $X$, can provide an accurate estimate of the next $y$. This possibility is noted as *posterior*. Also, $P(y|X, w)$ indicates the probability that $w$ corresponds to previous observations of $X$ and $y$ (training data). This probability is noted as *likelihood*. Also, $p(w)$ shows the probability of occurring $w$ in previous observations or the same training samples. This probability is hereinafter referred as the default probability.

In Gaussian process regression, first a default Gaussian process that can be determined by a covariance and mean function is considered. In particular, a Gaussian process is in form of a multivariate, unlimited, Gaussian distribution, in which each set of data labels have a common Gaussian distribution. In this default Gaussian process, we can combine prior knowledge of function space by selecting mean and covariance functions.

To estimate the target variable (posterior probability distribution), the test observation and data are conditioned. Given that a Gaussian process is selected as the distribution of the default probabilities, the distribution of the prediction variable can also be calculated and leads to a Gaussian distribution that is definable by the covariance and mean.

### 3.5. Weighting of estimation algorithms by cuckoo search

As mentioned earlier, after training the learning models, the optimal weight of each learning model is determined using the CS optimization. The computational steps of this optimization algorithm were described in Section 2-1. In the proposed system, the same mechanism is used for optimizing the weight values of learning models. Therefore, considering these computational steps, we will describe the structure of solution vector and the fitness function in this optimization algorithm.

The purpose of the CS optimization in the proposed system is to determine the optimal weight values for the weights $\langle a, b, c \rangle$ of the learning models in Equation (1). The cuckoo search algorithm determines these values in the range of $[0,1]$, in such a way that the learning models which have a higher error, will have a lower weight value; and the more accurate learning models will have a higher weight. With this explanation, the length of each solution in the optimization algorithm will be 3 and the search bound of all variables is equal to $[0,1]$.

Fitness function, is the key part of an optimization algorithm. A fitness function describes the optimality of a solution. As a result, using the fitness function, it can be determined which learning model output will have a higher value and which response found by the search algorithm is optimal. If the goal of the optimization algorithm is to determine the optimal weight for the output of each learning model, then a response vector is displayed as $A = \{a, b, c\}$, where $a$ represents the weight for the ANN output, $b$ refers to the weight specified for CART output and $c$ represents the weight specified for the output of GPR model. The fitness of a solution in cuckoo search is calculated as follows:

$$fitness = \frac{1}{1 + RMSE} \tag{10}$$

Where, RMSE refers to the Root Mean Squared Error:

$$RMSE = \sqrt{\frac{1}{n} \sum_{i=1}^{n} (O_i - \frac{a.W_i + b.X_i + c.Y_i}{a+b+c})^2} \tag{11}$$

In the above equation, $O_i$ refers to the real price of the stock. Also, $W_i$, $X_i$ and $Y_i$ are the outputs estimated by each of the prediction algorithms in the proposed method and $a$, $b$ and $c$ are the weight values for each of these prediction algorithms.

Thus, the cuckoo search algorithm tries to optimize the weight parameters of the learning models using the described fitness function. After determining the optimal values for the weight vector of $\langle a, b, c \rangle$, Equation (1) is used for determining the stock price for new test samples.

## 4. Results and discussion

In this section, the results of implementing the proposed method and testing it on a real database will be presented. The MATLAB software was used for implementation of the proposed prediction system. In the following, after stating the specifications of the database used, we will discuss about the implementation results.

The database used in the experiments has been collected through the Tehran Stock Exchange database [37]. This database includes the stock value of 70 companies in the Tehran Stock Exchange. The stock price records stored in this database cover the period from December 2018 to June 2022. According to the purpose of the research, the data of 10 construction companies (which fully cover the target period) have been used in these tests.

In the database used, contains 503 data records for each company. These records contain the stock price information of each company in the activity days of Tehran Stock Exchange market. 402 data samples of each company are used for the learning phase and training the prediction system and the rest for the test. Figure (4) shows a chart of stock price changes for one of the companies in the database.

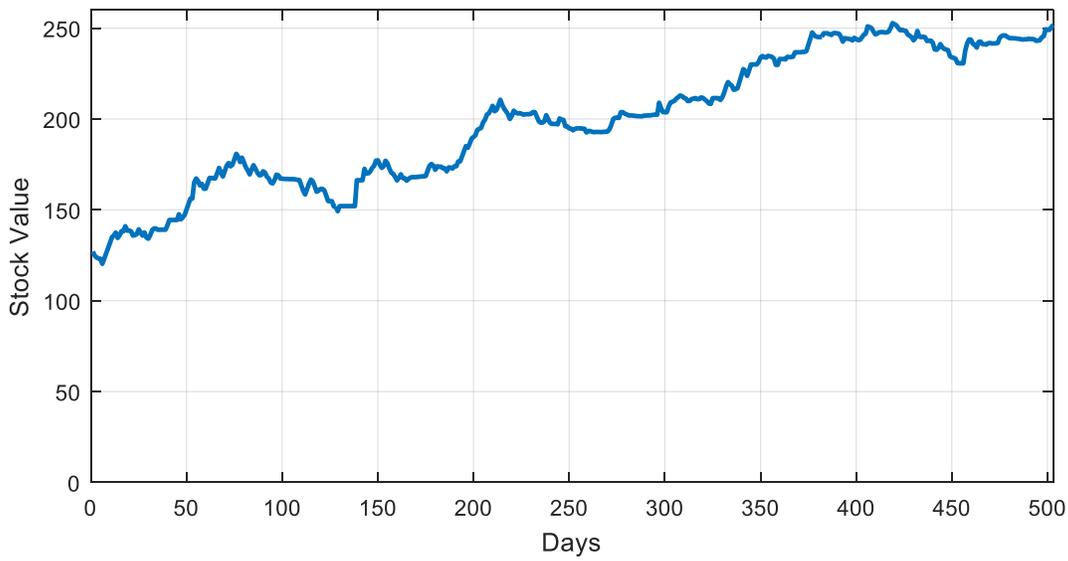

*Figure 4: Chart of stock price changes for one of the companies in the database*

The proposed method performs the prediction operation based on a combination of the results of three different algorithms. One of the algorithms used in the proposed method is an ANN with two hidden layers and Levenberg-Marquardt training function. Figure (5) compares the real and the predicted values by this ANN, for the stock price of the test samples.

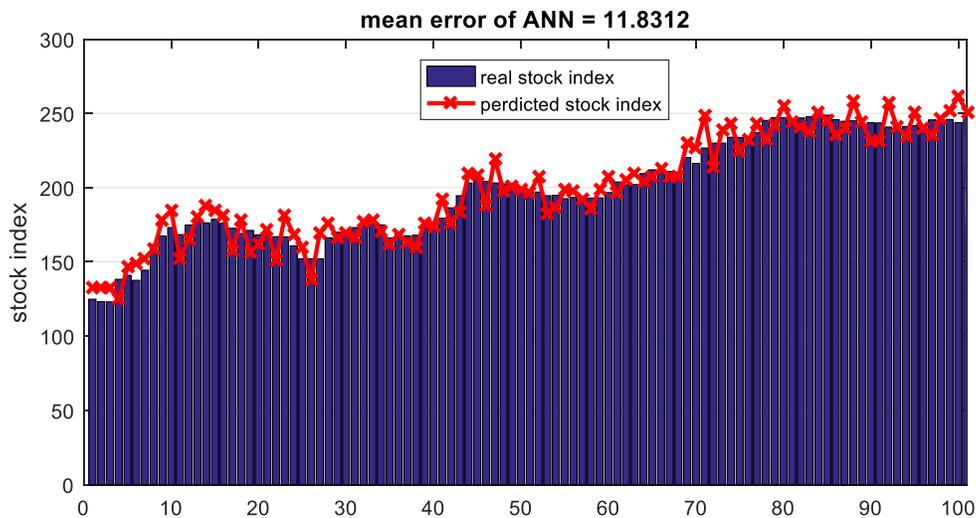

*Figure 5: ANN performance in predicting the stock price of the Kerman data*

These results indicate that the MAE of the used ANN for the Kerman data is 11.83. The second algorithm used in the proposed method is GPR. Figure (6) compares the real and predicted stock price values by this learner for the price values of the test samples.

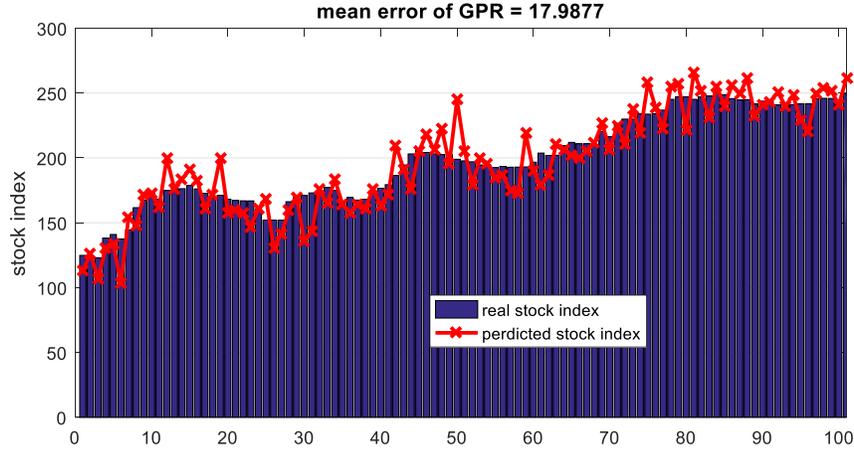

*Figure 6: GPR performance in predicting the stock price of the Kerman data*

These results show that the mean absolute error of GPR for the Kerman data is 17.99. The third prediction algorithm used in the proposed method is CART. Figure (7) illustrates the performance of CART in predicting Kerman stock price.

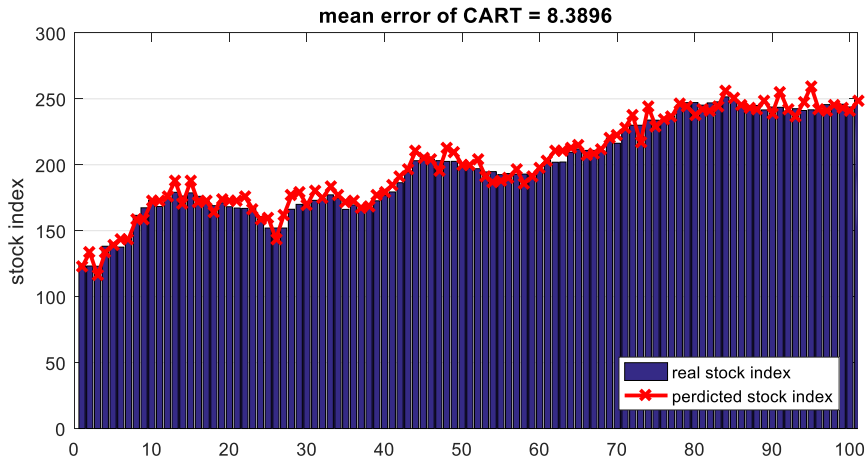

*Figure 7: CART performance in predicting the stock price of the Kerman data*

According to the results, the mean absolute error of the CART for the Kerman data is 8.39. As the test results of the three estimation algorithms show, the CART has the lowest error and the ANN is in the second place and finally GPR produces the highest error.

In the proposed approach, using the CS algorithm, the learning models were weighted and the impact of each algorithm on the final output was determined. Based on the results obtained; CART, ANN and GPR are weighted as 0.915, 0.876 and 0.131, respectively. Thus, if for a test sample, the ANN produces output $X_i$, CART outputs $Y_i$, and GPR produce output $Z_i$, then the final output of the proposed method is calculated as $\frac{0.876 \times X_i + 0.915 \times Y_i + 0.131 \times Z_i}{0.876 + 0.915 + 0.131}$.

Figure (8) illustrates the actual and predicted values using the proposed method (weighted ensemble learning) for the stock price of the test samples.

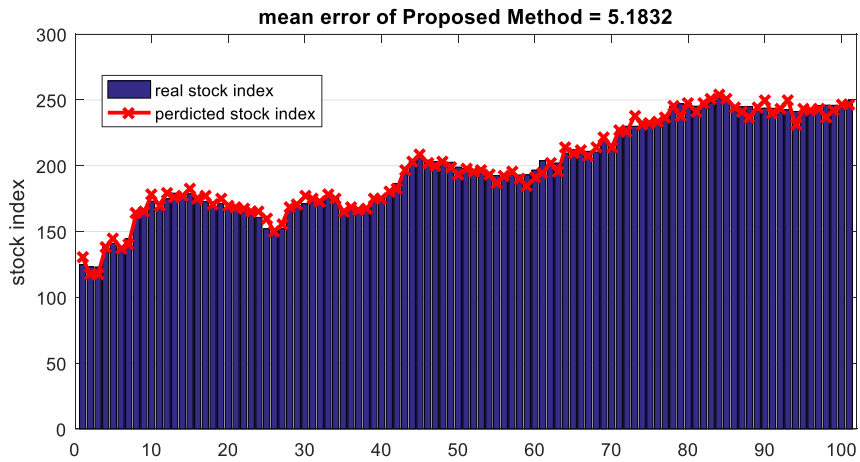

*Figure 8: Performance of the proposed method in predicting the stock price of the Kerman data*

According to the results, the mean absolute error of the proposed method for the Kerman data is 5.18. Therefore, the proposed system is able to effectively reduce the prediction error using weighted ensemble learning. Figure (9) compares the error rate obtained by the proposed system, with the algorithms used in its ensemble model for predicting the price of all 10 construction companies. It should be noted that the error rate criterion is calculated by dividing the output difference of the learning model (prediction error) by the real price of the stocks.

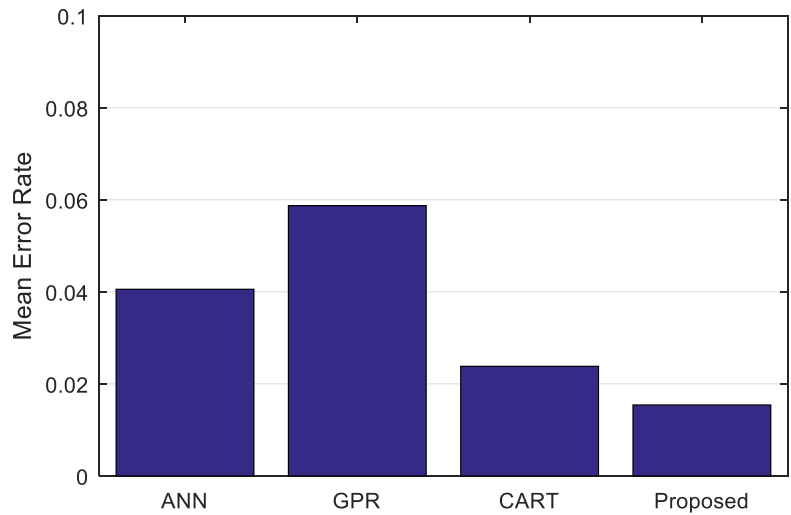

*Figure 9: Error rate of the proposed method and other algorithms in predicting price of construction companies*

According to these results, the proposed system can effectively reduce the error rate in predicting the price of stocks. Figure 10, illustrates the regression plot of the proposed method and its learning components. In figure 11, the error histogram plots are shown. According to Figure 10, the regression plot obtained from the prediction of the proposed system has lower difference with the actual values and the slope of the regression line is closer to the value of 1 with lower bias value, which indicates that the proposed system provides a more accurate prediction than the compared algorithms. On the other hand, comparing the error histograms of Figure 11, indicates that the error range of the proposed system is more limited than other methods. These characteristics confirm that using the weighted ensemble technique can be effective in increasing the reliability of the results.

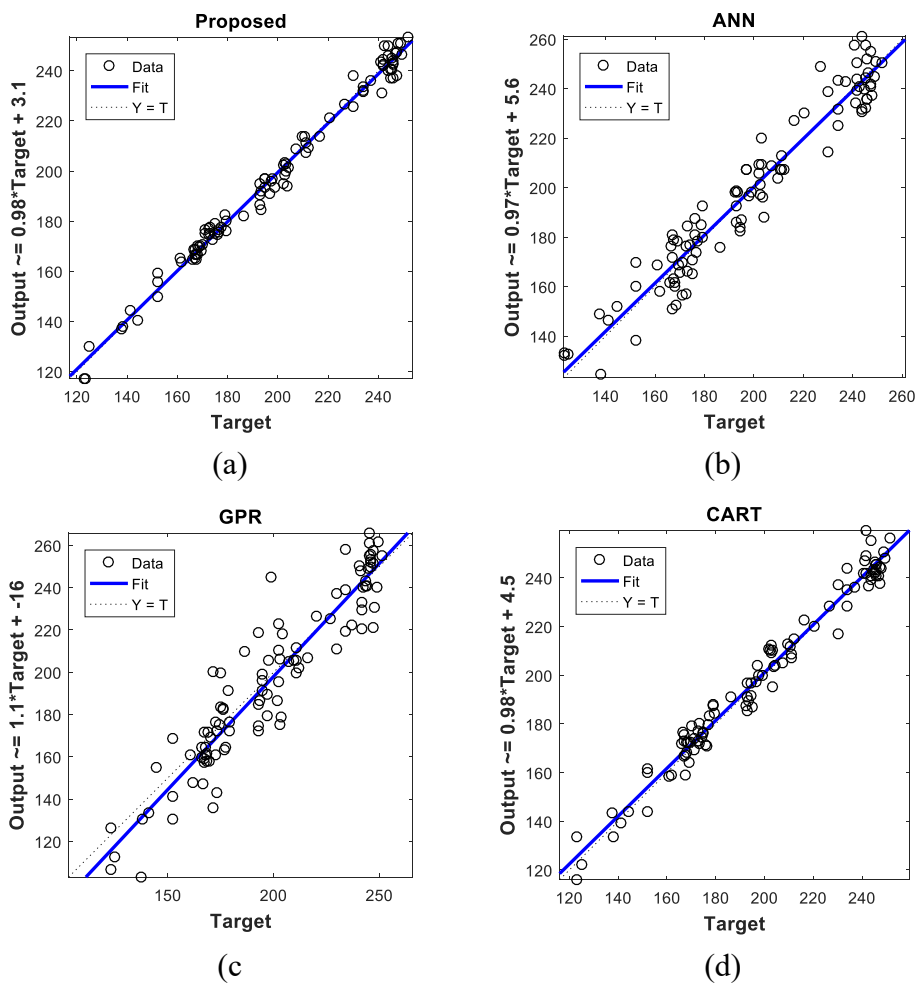

*Figure 10. Regression plots for (a) proposed method, (b) ANN, (c) GPR and (d) CART*

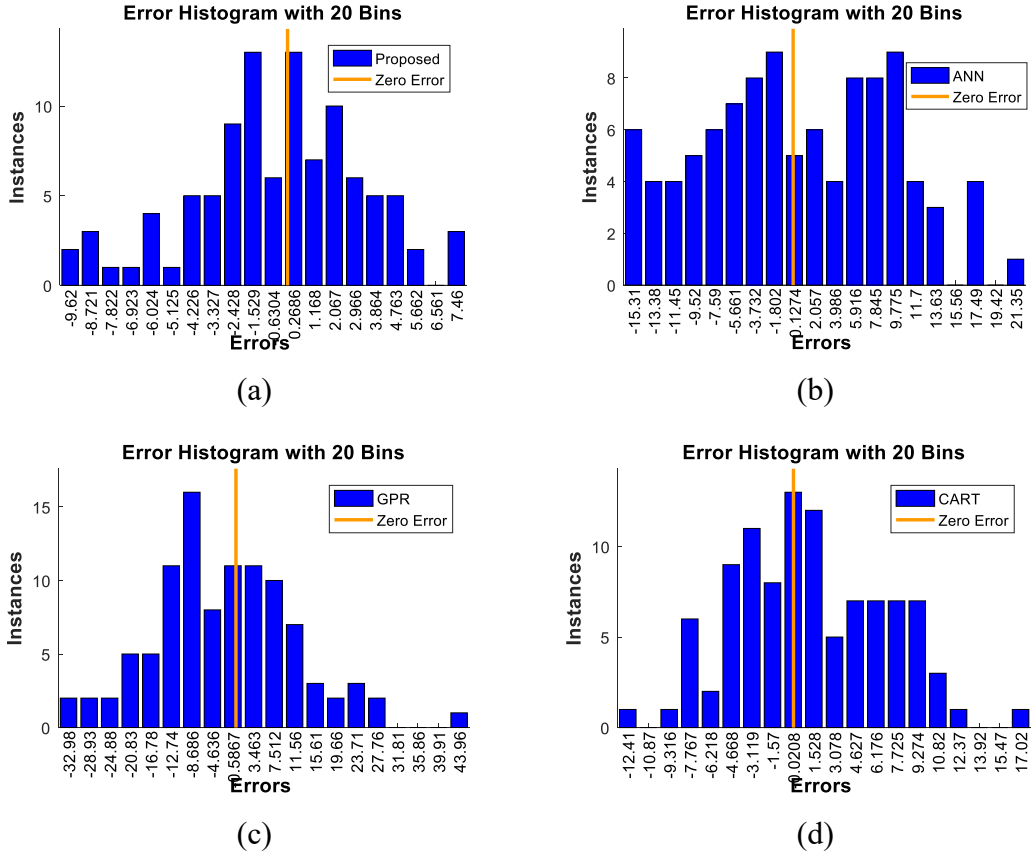

*Figure 11. Error histograms for (a) proposed method, (b) ANN, (c) GPR and (d) CART*

In Table (3), average prediction error rate of the proposed system is compared with other learning algorithms on the various companies' data.

**Table 3: Average prediction error rate of proposed method for one-day intervals**

| Stock | Proposed | GPR | ANN | CART |
|---|---|---|---|---|
| KERMAN | 0.0154 | 0.0587 | 0.0405 | 0.0238 |
| SSHAHED | 0.0365 | 0.0664 | 0.0504 | 0.0526 |
| SNOSA | 0.0226 | 0.0825 | 0.0606 | 0.0299 |
| SABAD | 0.0533 | 0.1730 | 0.0599 | 0.0745 |
| SROOD | 0.0080 | 0.0402 | 0.0241 | 0.0110 |
| SNOOR | 0.0796 | 0.3292 | 0.0810 | 0.1175 |
| SGHARB | 0.0591 | 0.1124 | 0.0686 | 0.0993 |
| SBEHSAZ | 0.0073 | 0.0153 | 0.0146 | 0.0131 |
| KISTOON | 0.0460 | 0.1321 | 0.0723 | 0.0759 |
| SOMID | 0.0372 | 0.1139 | 0.0533 | 0.0927 |

Table (4) also compares the MAE of the proposed system and other learning models in predicting the stock price of these companies.

Table 4: Mean absolute error of proposed method for one-day interval

| Stock | Proposed | GPR | ANN | CART |
|---|---|---|---|---|
| KERMAN | 5.1832 | 17.9877 | 11.8312 | 8.3896 |
| SSHAHED | 5.4107 | 8.7111 | 5.5386 | 6.0083 |
| SNOSA | 2.3769 | 8.6723 | 6.3693 | 3.1436 |
| SABAD | 5.6003 | 18.1723 | 6.2855 | 7.8164 |
| SROOD | 0.8442 | 4.2266 | 2.5331 | 1.1527 |
| SNOOR | 8.3578 | 34.5848 | 8.5004 | 12.3472 |
| SGHARB | 6.2169 | 11.7967 | 7.2084 | 10.4439 |
| SBEHSAZ | 0.7665 | 1.6101 | 1.5340 | 1.3839 |
| KISTOON | 4.8367 | 13.8799 | 7.5978 | 7.9697 |
| SOMID | 3.9185 | 11.9737 | 5.6022 | 9.7522 |

Tables (3) and (4), demonstrate the error rate of the proposed method for predicting stock price of each company in next day. According to these results, the proposed system in all cases can reduce the prediction error. Therefore, the use of weighted ensemble approach can be used as an efficient technique in improving the accuracy of prediction algorithms. According to these results, the average error rate of the proposed method for predicting the stock price of all companies is 0.036 and the MAE is 4.351. Achieving an average error rate of 0.036 means the efficiency of the proposed system in modeling stock behavior, and therefore the proposed method can be used as an efficient tool in real-world scenarios.

In order to better In order to better evaluate the efficiency of the proposed method, the prediction interval is increased from one day to one week. In this case, Equation (7) will be written in form of $p(k + 7) = f(s(k))$; which means the proposed ensemble system must predict the value of stock in next week. Also, the *n* and *t* parameters in Equation (2) have been set as $n = 5$ and $t = 7$, which means the stock prices of past five weeks are used as input sequence of the system. Table (5), compares the average prediction error rate of the proposed system in daily and weekly intervals. Table (6) also, compares the MAE of the proposed system and other learning models in daily and weekly intervals.

As shown in Tables (5) and (6), increasing prediction interval form daily to weekly period, results in a slight increase in prediction error of the proposed method. Nevertheless, the amount of error increase in the proposed ensemble system is insignificant compared to other learning models.

**Table 5: Comparing the average prediction error rate of the proposed system in daily and weekly intervals**

| Algorithm | average daily prediction error rate | average weekly prediction error rate |
|---|---|---|
| **Proposed** | **0.0365** | **0.0378** |
| GPR | 0.1124 | 0.1353 |
| ANN | 0.0525 | 0.0848 |
| CART | 0.0590 | 0.0997 |

**Table 6: Comparing the MAE of the proposed system in daily and weekly intervals**

| Algorithm | MAE for daily prediction | MAE for weekly prediction |
|---|---|---|
| **Proposed** | **4.3512** | **4.4895** |
| GPR | 13.1615 | 19.5667 |
| ANN | 6.3001 | 7.5210 |
| CART | 6.8407 | 8.1562 |

Examining the values of these tables, shows that each of the learning models individually experiences a greater increase in error criteria. This is despite the fact that the increase in the error criteria in the proposed method is much less. This reduction in error can be considered as the result of using the CS algorithm in determining the optimal weight values of each learning component. In the proposed method, the cuckoo search algorithm tries to determine the weight values of each learning model in such a way that the final error rate of the aggregate system is minimized. This condition is drawn in figure (12).

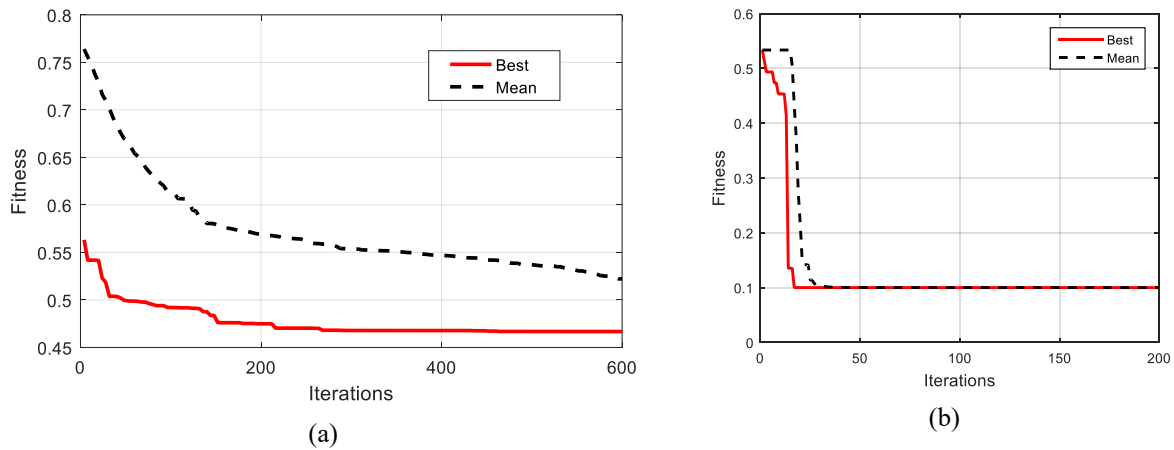

*Figure 12: Best/mean fitness values of the CS algorithm during each iteration of optimizing weight coefficient of learning components in (a) daily and (b) weekly prediction scenarios*

Based on the diagrams illustrated in Figure (12), the CS algorithm can minimize the prediction error (fitness) in the weighted ensemble system by efficiently searching the problem space and determining the optimal combination of weight values.

## 5. Managerial insights and practical implications

Providing a reliable view of upcoming changes in a company's assets is an important requirements in managing organizations. This issue is very important for companies of construction section, due to the dependence of their value on various political, economic, social, etc. situations. The stock value of companies in construction section can reflect the performance of those companies well. For this reason, this research focused on predicting the stock value of these companies; and a method based on weighted ensemble learning was presented, which is able to predict the stock value of construction companies in daily to weekly intervals. The ability to predict stock value on daily to weekly intervals means that the proposed system is an efficient method for short-term to mid-term prediction and can provide a reliable view of company performance. Achieving the average error rates of 0.0365 (in daily prediction) and 0.0378 (in weekly prediction), confirms that the proposed system is practically useful in two scenarios: First, this system can provide an accurate view of the upcoming financial changes which the organization may face, and inform managers of these possible changes, so that they can prevent critical situations from occurring. Second, the proposed method can be useful for investors of construction section and help them determine the right place for investment.

## 6. Conclusion

In this study, the effectiveness of the ensemble learning technique for predicting stock price of companies in construction section was investigated. In general, combining several learning models can be effective in increasing the accuracy of prediction systems. However, one of the most important points in design of ensemble systems is to determine the impact of each learning component on the final output of the system, which in the proposed method was done using the CS optimization. The performance of the proposed method was evaluated in two scenarios of daily and weekly prediction of stock price. According to the results:

- The proposed system was able to predict the daily changes in stock price with average error rate of 0.0365 and MAE of 4.3512. By comparing the results of the proposed system with other learning algorithms, we can conclude that our system has a higher accuracy in stock

- price prediction, which is the result of using optimal coefficients in weighted ensemble learning system.
- The proposed system could predict the one-week-ahead value of stock price with average error rate of 0.0378 and MAE of 4.4895. In this case, each of the learning models individually experiences a greater increase in error criteria. This is despite the fact that the increase in the error criteria in the proposed method is much less, which is the result of using the CS algorithm in determining the optimal weight values of each learning component.
- From the results it can be concluded that the proposed system can be practically usable in assisting managers of construction companies by providing an accurate view of upcoming financial changes; as well as helping the investors of construction section to determine the right place for investment.

One of the limitations of the current research is that in the proposed system, the prediction is performed only using extracted data from the index and statistical information of the stock market; while, in order to achieve better performance, other information such as: news, gold prices, oil prices, etc. can be used. Therefore, the use of these features in the proposed model can be studied in future work. Another limitation of the proposed method is the increase in its processing time. Although the ensemble technique can be effective in increasing the prediction accuracy, it increases the processing time due to the use of multiple learning models. However, it should be noted that this increase in processing time is noticeable only during training phase. The application of other optimization algorithms like Genetic Algorithm (GA) or Ant Colony Optimization (ACO) can be considered to determine the optimal weights of learners. Also, using deep learning approaches in the proposed ensemble model may be effective in achieving a more efficient predictive system.